# Generation of transversely oriented optical polarization Möbius strips

LIXIU SU,[1] XINDONG MENG,[1] YU XIAO,[1] CHENHAO WAN,[1,2,*] AND QIWEN ZHAN[2,*]

[1]*School of Optical and Electronic Information and Wuhan National Laboratory for Optoelectronics, Huazhong University of Science and Technology, Wuhan, Hubei 430074, China*
[2]*School of Optical-Electrical and Computer Engineering, University of Shanghai for Science and Technology, Shanghai 200093, China*
*\* wanchenhao@hotmail.com, qwzhan@usst.edu.cn*

**Abstract:** We report a time-reversal method based on the Richards-Wolf vectorial diffraction theory to generate transversely oriented optical Möbius strips that wander around an axis perpendicular to the beam propagation direction. A number of sets of dipole antennae are purposefully positioned on a prescribed trajectory in the $y = 0$ plane and the radiation fields are collected by one high-NA objective lens. By sending the complex conjugate of the radiation fields in a time-reversed manner, the focal fields are calculated and the optical polarization topology on the trajectory can be tailored to form prescribed Möbius strips. The method can be extended to construct various polarization topologies on three-dimensional trajectories in the focal region. The ability to control optical polarization topologies may find applications in nanofabrication, quantum communications, and light-matter interactions.

## 1. Introduction

The engineering of structured light enables full control of the spatial and temporal features of optical fields [1]. The combination of tailoring amplitude, phase, and state of polarization creates optical topological structures commonly related to singular optics, such as phase singularities and polarization singularities [2-7]. Topological structures including Möbius strips, ribbons, and knots are more than fancy optical complexity, but has found applications in quantum communications as information carriers [8-17].

Optical Möbius strips, featuring a one-sided surface, are formed by the major axes of three-dimensional polarization ellipses. Optical Möbius strips can be created by tightly focusing optical fields emerging from a *q*-plate [10]. The interference of two noncoaxial circularly polarized beams of opposite handedness with different scalar topological charges also result in Möbius strips topological structures [9,18]. Möbius strips are also discovered in light scattered from high-index dielectric nanoparticles [19].

Most optical Mobius strips discovered so far are within the cross section of a beam. In the current paper, we report a time-reversal method based on the Richards-Wolf vectorial diffraction theory for generating transversely oriented Möbius strips that wander around an axis perpendicular to the beam propagation direction [20-27]. A number of dipole antennae are positioned on a circular trajectory in the *x-z* plane and their radiation fields are collected by one high-NA microscope objective. By sending the radiation fields in time reversed order, optimal light spot arrays are formed at the source location, forming transversely oriented ring-shaped focal fields with Möbius strips polarization topologies. The method can be further extended to create other optical topological structures along arbitrary prescribed trajectories in three-dimensional space. The ability to tailor optical polarization topologies may spur various applications in nanofabrication and light-matter interactions.

## 2. Time reversal method

The time-reversal theory indicates that an optimal light spot can be formed at the source location by sending in time reversed order the radiation fields received from an infinitesimal dipole antenna. If three dipole antennae with orthogonal oscillating directions are positioned at the same location, not only can an optimal light spot be formed but also its three-dimensional state of polarization can be controlled by the relative amplitude and phase of the individual dipole antennae. In this work, we demonstrate that if a number of sets of dipole antennae are positioned on a prescribed trajectory, it is even possible to tailor the optical polarization topologies along arbitrary three-dimensional curves in a time-reversed manner.

Figure 1 shows the schematic of the proposed time-reversal scheme. To reconstruct the polarization topology of Möbius strips, a number of sets of dipole antennae are uniformly positioned in the $y = 0$ plane along a circular trajectory of radius $r_0$. Each set of dipole antennae occupy one spot on the trajectory and contain three individual dipoles oscillating along the $x$, $y$, and $z$ directions. The radiation fields generated from all dipole antennae are coherently combined in the curved surface of a high-NA objective lens.

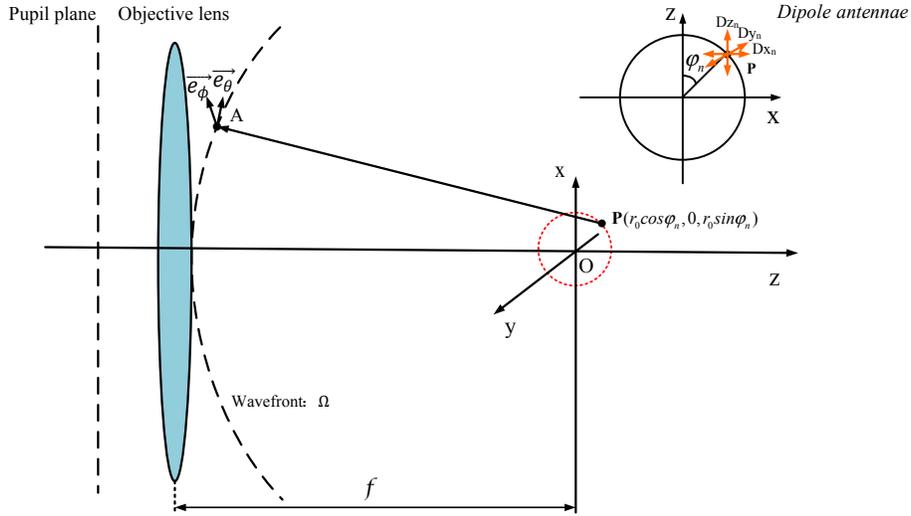

**Fig. 1. Schematic of the proposed time-reversal method.** Dipole antenna groups are uniformly positioned in the x-z plane along a circular trajectory of radius $r_0$. The radiation fields generated from all dipole antennae are collected by one high-NA objective lens.

The location of each set of dipole antennae is expressed as $(r_0 \sin\varphi_n, 0, r_0 \cos\varphi_n)$, where $\varphi_n \in [0, 2\pi)$ is the angle with respect to the $z$ axis in the $x$-$z$ plane. Based on the antenna radiation theory [23], the optical fields in the wavefront surface $\Omega$ of the objective lens are given by:

$$\overrightarrow{E_\Omega}(\theta,\phi) = \sum_{n=1}^{N} \left[ \overrightarrow{E_{xn}} + \overrightarrow{E_{yn}} + \overrightarrow{E_{zn}} \right]$$

$$= \sum_{n=1}^{N} j\eta \frac{kI_0 \exp(-jkr)}{4\pi r} \left[ \tilde{A}_{xn}\left(-\cos\theta\cos\phi \overrightarrow{e_\theta} + \sin\phi \overrightarrow{e_\phi}\right) + \tilde{A}_{yn}\left(-\cos\theta\sin\phi \overrightarrow{e_\theta} - \cos\phi \overrightarrow{e_\phi}\right) + \tilde{A}_{zn}\sin\theta \overrightarrow{e_\theta} \right]$$

$$\cdot \exp\left[ jkr_0\left(\cos\varphi_n \cos\theta - \sin\varphi_n \sin\theta\cos\phi\right)\right],$$

(1)

$\overrightarrow{E_{xn}}$, $\overrightarrow{E_{yn}}$ and $\overrightarrow{E_{zn}}$ represent the radiation fields of the *n*th dipole oscillating along the $x$, $y$ and $z$ directions, and $N$ is the number of sets of dipole antennae. $\tilde{A}_{xn}$, $\tilde{A}_{yn}$ and $\tilde{A}_{zn}$ represent the relative complex amplitude of the *x*, *y*, *z* dipole antennae of the *nth* set. $\vec{e}_\theta$ and $\vec{e}_\phi$ are unit vectors along the elevation and azimuthal directions in the curved surface. It is noted that a *z* dipole generates only the elevation component. $\eta$ is the impedance, $k = 2\pi/\lambda$ is the wave number, $I_0$ is the constant electric current, $\theta$ is the elevation angle, $\phi$ is the azimuthal angle, and *r* is radius of curvature of the wavefront $\Omega$. The path length difference between OA and PA is given by $r_0(\cos\varphi_n \cos\theta - \sin\varphi_n \sin\theta \cos\phi)$.

The objective lens obeys the *sine* condition $r = f\sin\theta$, where *f* is the focal length, therefore the projection function is given by $\vec{E}_i(r,\phi) = \vec{E}_\Omega(\theta,\phi)/\sqrt{\cos\theta}$, where $\theta = \sin^{-1}(r/f)$ [24].

The electric field data of Möbius strip polarization topologies is obtained from a tightly focused beam emerging from a *q*-plate based on the Richards-Wolf vectorial diffraction theory. The *x*-, *y*-, and *z*-components of the electric field data at each sampled point are connected with the relative amplitude and phase of the three orthogonally oscillating dipoles of each set of dipole antenna on the prescribed trajectory. The complex conjugate of the radiation fields received from all sets of dipole antennae are focused back with one high-NA objective lens. According to the Richards-Wolf vectorial diffraction method, the electric fields in the focal region are given by [26]:

$$\vec{E}(r_p, \Psi, z_p) = \frac{jk}{2\pi} \int_0^{\theta_{max}} \int_0^{2\pi} \vec{E}_\Omega(\theta,\phi) \exp(jkr_p \sin\theta \cos(\phi - \Psi) + jkz_p \cos\theta) \sin\theta \, d\theta \, d\phi, \qquad (2)$$

where $\vec{E}_\Omega(\theta,\phi)$ are the electric fields in the wavefront surface, $(r_p, \Psi, z_p)$ is the cylindrical coordinate in the focal region, $\theta_{max} = \sin^{-1}(NA)$ and NA is the numerical aperture of the objective lens.

We examine the three-dimensional electric fields and the associated three-dimensional states of polarization in the *y* = 0 plane in the focal region. The optical polarization topology, traced along a circular trajectory, can be deciphered by deriving the major and minor axes and the normal vector of the three-dimensional polarization ellipses [10].

$$\vec{\alpha} = \frac{1}{\sqrt{\vec{E}\cdot\vec{E}}} \mathrm{Re}\left(\vec{E}^* \cdot \sqrt{\vec{E}\cdot\vec{E}}\right)$$
$$\vec{\beta} = \frac{1}{\sqrt{\vec{E}\cdot\vec{E}}} \mathrm{Im}\left(\vec{E}^* \cdot \sqrt{\vec{E}\cdot\vec{E}}\right) \qquad (3)$$
$$\vec{\gamma} = \mathrm{Im}\left(\vec{E}^* \times \vec{E}\right),$$

where *Re*, *Im* and $\vec{E}^*$ denote the real part, imaginary part, and the complex conjugate of $\vec{E}$, respectively.

### 3. Numerical simulation

Two examples are considered to demonstrate that the method is capable of generating transversely oriented optical polarization Möbius strips in the *y* = 0 plane in the focal region. The parameters used in the numerical simulation include: the number of sets of dipole antennae *N* = 12, the numerical aperture of the objective lens NA = 1, and the radius of the circular trajectory $r_0 = \lambda$.

For the first example, the data of optical fields generated from a *q*-plate of topological charge 1/2 is evenly sampled and interpolated. The *x*-, *y*-, and *z*-components of the electric field data is assigned to the *x*-, *y*-, and *z*-dipole at each point on the circular trajectory of radius $r_0 = \lambda$. The optical fields in the pupil plane are calculated based on the projection theory and the intensity and polarization distributions are displayed in fig. 2(a). The intensity distributions of $E_x$ and $E_y$ components are shown in figs. 2(b) and 2(c). The phase distributions of $E_x$ and $E_y$ components are shown in figs. 2(d) and 2(e), respectively.

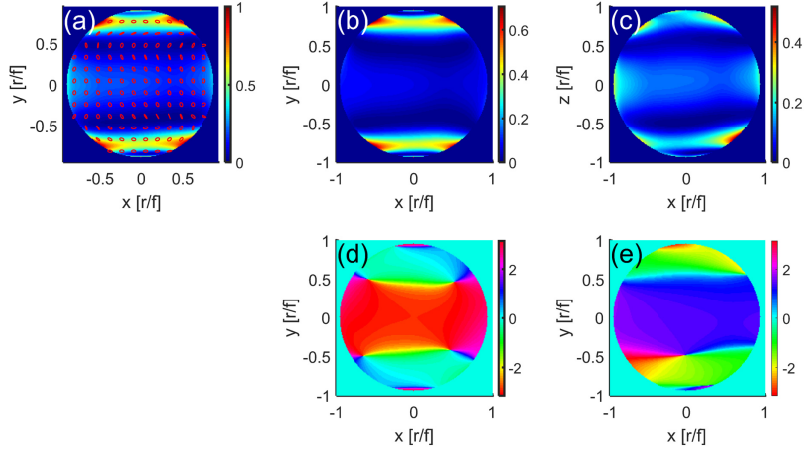

**Fig. 2.** (a) The intensity and polarization distributions in the pupil plane; (b)-(c) The intensity distributions of $E_x$ and $E_y$ components in the pupil plane, respectively; (d)-(e) The phase distributions of $E_x$ and $E_y$ components in the pupil plane, respectively.

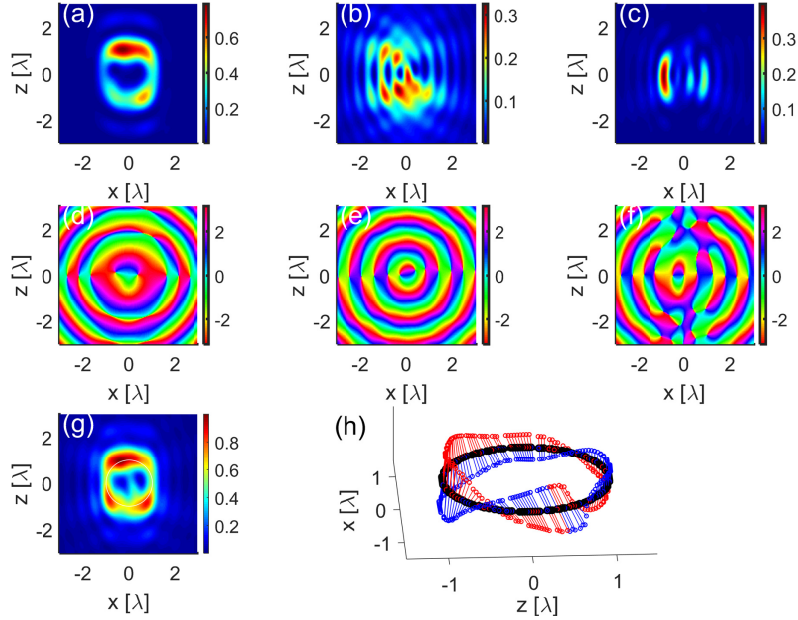

**Fig. 3. A Möbius strip with topological charge of -1/2 in the $y = 0$ plane.** (a)-(c) The intensity distributions of $E_x$, $E_y$ and $E_z$ components in the $y = 0$ plane, respectively; (d)-(f) The phase distributions of $E_x$, $E_y$ and $E_z$ components in the $y = 0$ plane, respectively; (g) The total intensity distributions in the $y = 0$ plane; (h) The major axes of three-dimensional polarization ellipses along the circular trajectory marked in 3(g) form a Möbius strip. The two halves of the major axes of the polarization ellipses are colored blue and red to indicate the orientation.

To determine the major axes of the polarization ellipses, it is necessary to calculate the amplitude and phase distributions of the complex vectorial optical fields in the desired plane. Figures 3(a) to 3(c) show the intensity distributions of the $E_x$, $E_y$ and $E_z$ components in the $y = 0$ plane, respectively. Figures 3(d) to 3(f) show the phase distributions of the $E_x$, $E_y$ and $E_z$ components in the $y = 0$ plane, respectively. The total intensity distributions are plotted in fig. 3(g), that is nearly ring-shaped as designed. The major axes of three-dimensional polarization ellipses form a Möbius strip polarization topology with topological charge of -1/2 as plotted in fig. 3(h). The two halves of the major axes of the polarization ellipses are colored blue and red to indicate the orientation.

The second example is to construct a Möbius strip with topological charge -3/2 in the $y = 0$ plane, that has five half-twists along a circle. The complex optical fields in the pupil plane are spatially variant in both intensity and polarization as shown in fig. 4(a). The intensity distributions of $E_x$ and $E_y$ components are displayed in figs. 4(b) to 4(c). The phase distributions of $E_x$ and $E_y$ components are shown in figs. 4(d) to 4(e), respectively.

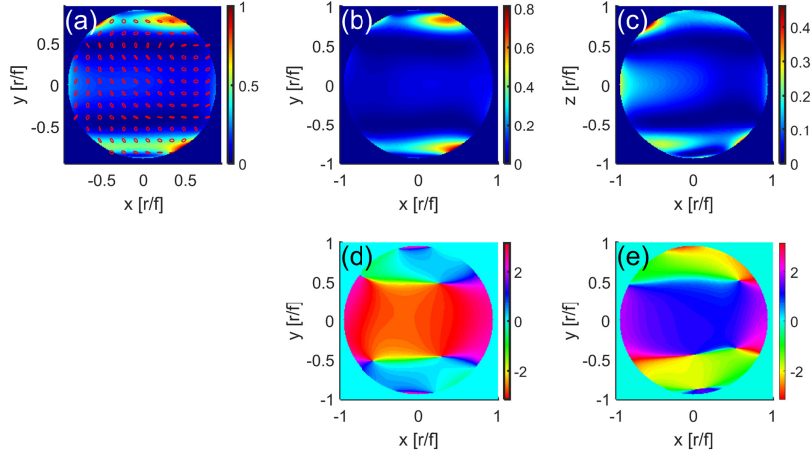

**Fig. 4.** (a) The intensity and polarization distributions in the pupil plane; (b)-(c) The intensity distributions of $E_x$ and $E_y$ components in the pupil plane, respectively; (d)-(e) The phase distributions of $E_x$ and $E_y$ components in the pupil plane, respectively.

Figures 5(a) to 5(c) plot the intensity distributions of the $E_x$, $E_y$ and $E_z$ components in the $y = 0$ plane, respectively. The phase distributions of the $E_x$, $E_y$ and $E_z$ components in the $y = 0$ plane are shown in figs. 5(d) to 5(f). The total intensity distributions are displayed in fig. 5(g). The major axes of three-dimensional polarization ellipses form a Möbius strip with topological charge of -3/2 as plotted in fig. 3(h). The generated Möbius strip polarization topology contains 5/2 twists as expected.

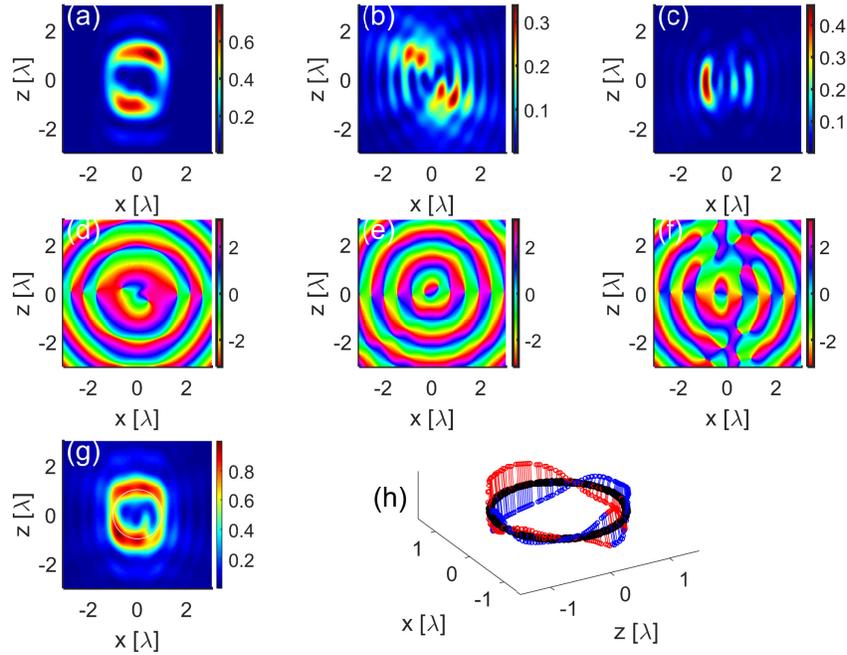

**Fig. 5. A Möbius strip with topological charge of -3/2 in the $y = 0$ plane.** (a)-(c) The intensity distributions of $E_x$, $E_y$ and $E_z$ components in the $y = 0$ plane, respectively; (d)-(e) The phase distributions of $E_x$, $E_y$ and $E_z$ components in the $y = 0$ plane, respectively; (g) The total intensity distributions the $y = 0$ plane; (f) The major axes of three-dimensional polarization ellipses along the circular trajectory marked in 5(g) form a Möbius strip. The two halves of the major axes of the polarization ellipses are colored blue and red to indicate the orientation.

In the next step, the behavior of the optical polarization topologies with different radii in the $y = 0$ plane and the optical polarization topologies in different $x$-$z$ planes near the focus are studied and compared. The Möbius strips with polarization topological charge of -1/2 along a circular trajectory of radius $r = \lambda$ in the $y = 0$ plane is displayed in fig. 6(a) as a reference. Figures 6(b) to 6(c) plot the polarization topologies along a circular trajectory of radius $0.9\lambda$ and $1.1\lambda$ in the $y = 0$ plane, respectively. Figures 6(d) to 6(e) plot the polarization topologies along a circular trajectory of radius $\lambda$ in the $y = -0.1\lambda$ plane and the $y = 0.1\lambda$ plane, respectively. Although the Möbius strips undergo some deformation, the topological charge remains unchanged for all cases.

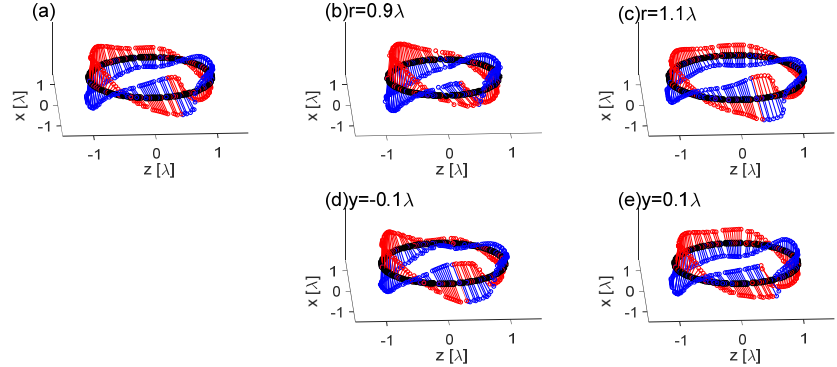

**Fig. 6. Möbius strips of different radii in various *x-z* planes.** (a) The polarization topologies along a circular trajectory of radius $r_0 = \lambda$ in the $y = 0$ plane; (b)-(c) The polarization topologies along a circular trajectory of radius $r_0 = 0.9\lambda$ and $r_0 = 1.1\lambda$ in the $y = 0$ plane, respectively; (d)-(e) The polarization topologies along a circular trajectory of radius $r_0 = \lambda$ in the $y = -0.1\lambda$ plane and the $y = -0.1\lambda$ plane, respectively. Two halves of the major axes of the polarization ellipses are colored blue and red to indicate the orientation.

## 4. Conclusions

In summary, we report a scheme that utilize the time-reversal method and the vectorial diffraction theory to generate transversely oriented optical Möbius strips polarization topology through tight focusing. The topological structures possessing polarization topological charges of -1/2 and -3/2 are constructed along prescribed trajectories in the *y* = 0 plane. The method can be further extended to build various polarization topologies along arbitrary three-dimensional trajectories in the focal region. Tailoring polarization topologies empowers applications in nanofabrication and quantum communications.

## Funding

## Disclosures

The authors declare no conflicts of interest.